\documentclass[letter,traditabstract]{aa} 
\usepackage{graphicx}
\usepackage{natbib}
\bibpunct{(}{)}{;}{a}{}{,} 

\usepackage{txfonts}
\begin{document}

   \title{The extent of dust in NGC~891 from Herschel/SPIRE images\thanks{{\it Herschel} is an ESA space observatory with science 
instruments provided by European-led Principal Investigator consortia and 
with important participation from NASA.}}

   \author{S. Bianchi\inst{1}
          \and
          E. M. Xilouris\inst{2}
          }
   \institute{INAF-Osservatorio Astrofisico di Arcetri, Largo E. Fermi 5, I-50125, Florence, Italy
         \and
             Institute of Astronomy and Astrophysics, National Obs. of Athens, 
             I. Metaxa and Vas. Pavlou, P. Penteli, GR-15236 Athens, Greece\\
              \email{sbianchi@arcetri.astro.it,xilouris@astro.noa.gr}
             }

   \date{}

 
  \abstract
{
We analyse Herschel/SPIRE images of the edge-on spiral galaxy NGC~891 at 250, 350, and 500~$\mu$m. 
Using a 3D radiative transfer model we confirm that the dust has radial fall-off similar to 
the stellar disk. The dust disk shows a break at about 12 kpc from the centre, where the profile
becomes steeper. Beyond this break, emission can be traced up to 90\% of the optical disk 
on the NE side. To the SW, we confirm dust emission associated with the extended, asymmetric HI disk,
previously detected by the Infrared Space Observatory (ISO). 
This emission is marginally consistent with the large
diffuse dust disk inferred from radiative transfer fits to optical images. 
No excess emission is found above the plane beyond that of the thin, unresolved, disk.
}


   \maketitle
%

\section{Introduction}

The dust distribution in the edge-on galaxy NGC~891 has been 
studied in detail. Radiative transfer fits to optical/near-infrared 
(NIR) images have shown that its extinction lane can be described
with a dust disk that is radially wider but vertically thinner than
the stellar disk. (Because its exponential radial scalelength is 1.5-2 
times and the vertical scalelength half that of the stars, so hereafter we
call it the {\em diffuse dust disk}; \citealt{XilourisA&A1998,XilourisSub1998}.)
This disk is found to have moderate extinction properties, since the central 
optical depth perpendicular to the disk is about 1 in the B-band. 
However, this is not large enough to
explain the amount of energy emitted by dust and observed in the 
far-infrared (FIR) and submm spectral energy distribution (SED):
the {\em energy balance} requires that up to three times more
starlight is absorbed by dust than predicted by the
{\em diffuse dust disk} \citep{PopescuA&A2000}. Similar dust disk 
properties and the same {\em energy balance} problem have been 
found in other edge-on galaxies 
\citep{XilourisSub1998,MisiriotisA&A2001,DasyraA&A2005,BianchiA&A2007,
BianchiA&A2008,BaesA&A2010b}.

Radiative transfer models of dust emission in NGC~891 solve the 
{\em energy balance} problem by including a second dust distribution 
associated with the molecular gas, of mass comparable to that of
the {\em diffuse dust disk}. This second component escapes 
detection in optical images because 
it is thinner than {\em diffuse dust} \citep{PopescuA&A2000,PopescuA&A2011}
or it is clumpy in nature, thus producing effects
that are less coherent than the extinction lane caused
by {\em diffuse dust} \citep{BianchiA&A2008}. Hereafter, we call
this second component the {\em clumpy dust disk}. A model for
NGC~891 was presented in \citet{BianchiA&A2008}, made
with the radiative transfer code TRADING. Though the model
includes several features, we found that emission beyond 100$\mu$m
is caused mainly by an exponential disk of evolved stars (radial scalelength 
4 kpc and vertical scalelength 0.4 kpc, as derived in the NIR by 
\citealt{XilourisSub1998}) heating at thermal equilibrium both
the {\em diffuse dust disk} (radial scalelength 8 kpc and vertical 
scalelength 0.2 kpc) and the external envelope of the clouds in the
{\em clumpy dust disk} (with clouds distributed exponentially with
the radial scalelength of the stars, and the vertical
scalelength of the diffuse dust disk). 

In this Letter we compare the structural properties of
the dust distribution in NGC~891 inferred indirectly from
extinction studies and/or model predictions with those 
derived from recent submm images taken by the SPIRE instrument 
\citep{GriffinA&A2010} aboard the Herschel Space Observatory 
\citep{PilbrattA&A2010}. In particular, the high sensitivity of 
SPIRE, together with the edge-on configuration of the galaxy,
allow studying submm emission up to large radial distances from
the galactic centre. 
This is needed to directly verify the presence of the diffuse 
dust disk, whose predicted emission could become visible 
over that of the clumpy dust disk only in the outskirts
of the galaxy.

\begin{figure*}
\sidecaption
\includegraphics[width=3.9cm]{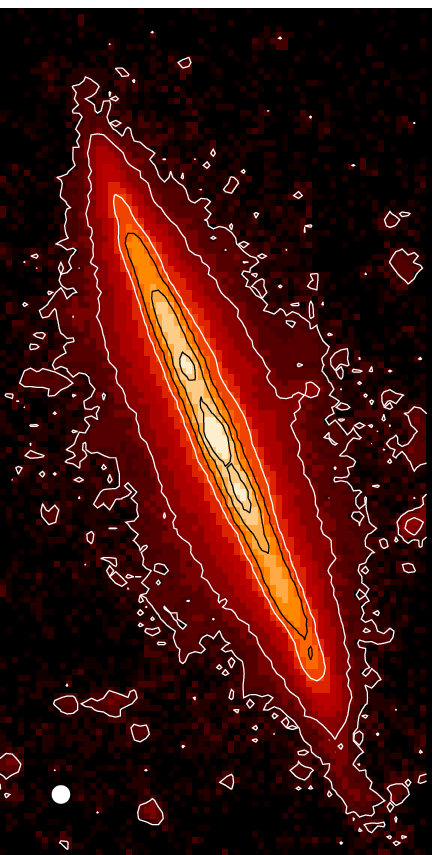} \, 
\includegraphics[width=3.9cm]{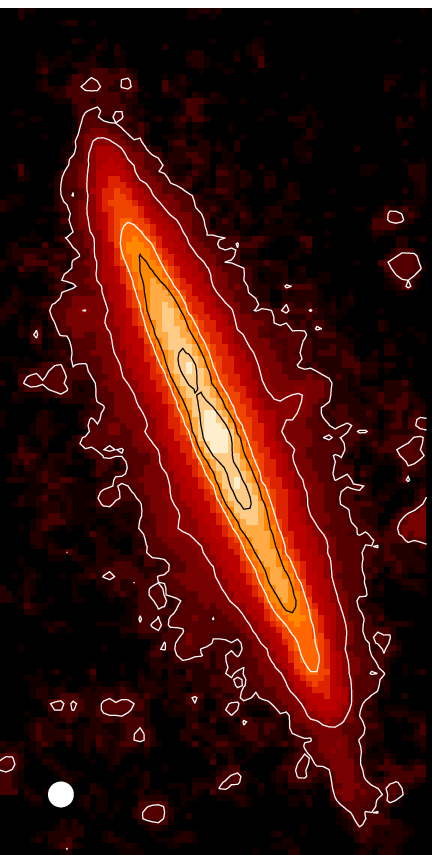} \,
\includegraphics[width=3.9cm]{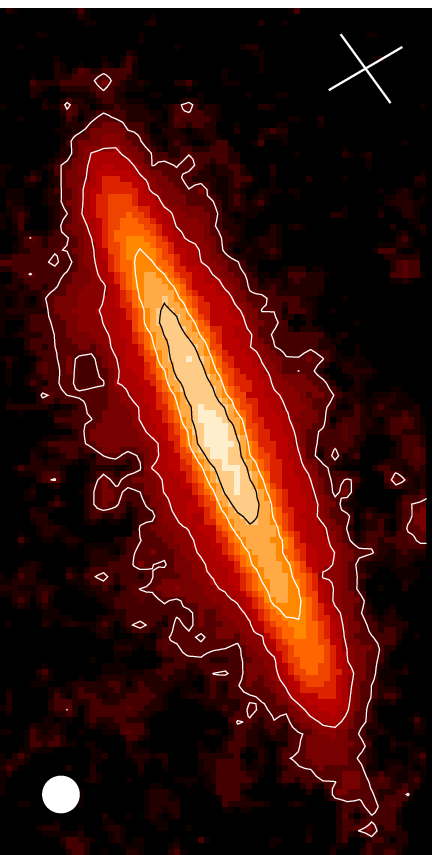}
\caption{SPIRE images of NGC~891 at 250 (left), 350 (centre), 
500$\mu$m (right). Each panel shows an area of 14$\arcmin\times$28$\arcmin$ 
centred on the galaxy's radio continuum coordinates
$\alpha = 2^\mathrm{h}\ 22^\mathrm{m}\ 33\fs0$, 
$\delta = 42\degr\ 20\arcmin\ 57\farcs2$
(J2000.0; \citealt{OosterlooAJ2007}).
The main beam size is indicated by the filled circle 
(FWHM: 18$\farcs$1, 24$\farcs$9, and 36$\farcs$4).
Images have pixel sizes of 6, 8, 12$\arcsec$ at 250, 350, and 500 $\mu$m, respectively
(i.e. about 1/3 of the measured FWHM).
The crossed lines in the 500$\mu$m image show the scan directions.
The sky rms noise is $\sigma=$ 0.74, 0.44, and 0.23 MJy sr$^{-1}$, from 250 to 500 $\mu$m. 
Contours are shown at 3, 10, 100, 200-$\sigma$ over all images, at 
500-$\sigma$ in the 250 and 350$\mu$m map, at 1000-$\sigma$ only in the
250$\mu$m map. North is up, east to the left.}
\label{figmap}
\end{figure*}

\section{Observations and data reduction}

SPIRE photometric observations at 250, 350, and 500 $\mu$m
were obtained as part of the Guaranteed Time Key Project Very Nearby Galaxy 
Survey (P.I. C. Wilson). The galaxy was observed
in large map mode, covering an area of 20$\arcmin$x20$\arcmin$ centred on the 
object with two crossscans, using a 30$\arcsec$ s$^{-1}$ scan rate.
Data was retrieved from the Herschel Science Archive and reduced
with the dedicate software HIPE\footnote{We used HIPE 5.2
and the SPIRE calibration tree v. 5.1. SPIRE
characteristics (point spread function - PSF - FWHM width and beam area, 
empirical PSF's, calibration uncertainty, and colour corrections) are taken 
from the SPIRE Observers' Manual (v2.3, 2011).}
\citep{OttProc2010}.  As the current pipeline 
simply removes median baselines from the timelines, causing
artifacts in the maps from the presence of bright extended sources,
we subtracted from each bolometer timeline a linear fit obtained
{\em after} carefully masking the data over the region covered by 
the galaxy. Maps were then produced using the na\"ive mapmaking 
procedure within HIPE.
The resulting images (Fig.~\ref{figmap}) have uniform backgrounds.
We also experimented with median baseline removal
and the alternative mapmaking software Scanamorphos v7
\citep{RousselA&A2011}, but found no significant difference for 
any of the analyses presented here. 

\section{The data vs the model}

Total flux densities are 169$\pm$25, 72$\pm$11, and 26$\pm$4 Jy, at 
250, 350, and 500 $\mu$m, respectively. These values include a 
colour correction appropriate to the observed spectral index 
($F_\nu \propto \nu^{2.7}$): since the source is only resolved along the
major axis (see later), we used a correction halfway between that for point 
and extended sources. (We multiplied the pipeline flux densities by 0.94, 
0.95, 0.94.)
{We adopted a very conservative 15\%
calibration uncertainty for extended emission (see the Observers' Manual, 
Sect. 5.2.13). Other sources of uncertainty (photometry, point source/extended
calibration) are much smaller, $\la 4\%$. }

\begin{figure}
\centering
\resizebox{\hsize}{!}{\includegraphics[trim = 0cm 0.5cm 0cm 0.2cm]{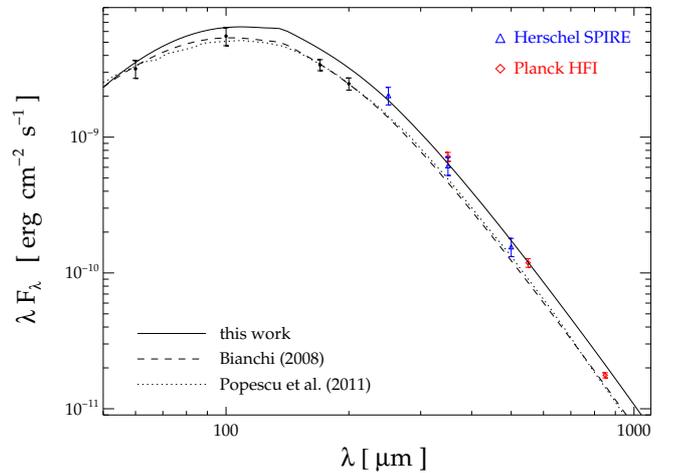}}
\caption{The integrated FIR-submm SED of NGC~891.
}
\label{figsed}
\end{figure}

SPIRE flux densities are shown in Fig.~\ref{figsed}, together with the
FIR data at 60 and 100$\mu$m from the IRAS satellite and at 170 and 
200$\mu$m from the ISO satellite \citep[see][and references therein]{PopescuA&A2011},
as well as the submm flux densities at 350, 550, and 850$\mu$m from the HFI 
instrument aboard the Planck satellite\footnote{Since NGC~891 is 
marginally resolved by Planck, we used the flux densities determined by 
fitting the data with a 2-D Gaussian model, as reported in the Early 
Release Compact Source Catalogue \citep{AdeA&A2011}. 
The quoted calibration uncertainties are 7\% at 350 and 
550$\mu$m and 2\% at 850 $\mu$m \citep{AdeA&A2011b}.  Colour corrections for 
a $\nu^{2.7}$ spectrum were taken from the Catalogue Explanatory Supplement.
We did not correct the flux density at 850$\mu$m for the
contamination due to the CO(3-2) line. This was found to
be about 5\% in SCUBA observations \citep{IsraelA&A1999}, and it is 
likely to be lower for the wider 353GHz bandpass. 
{The Planck flux density at 350$\mu$m appears to be higher than SPIRE, though 
still within 1-$\sigma$, for the large SPIRE calibration uncertainty adopted here}. 
We do not use other submm observations from
the literature \citep[for references, see:][]{PopescuA&A2011,BianchiA&A2008}, 
because they could have been maimed by a limited spatial coverage and
uncertain PSF sidelobe corrections. We note, however, that
the Planck flux density at 850$\mu$m is within the errors of SCUBA observations 
\citep{AltonApJL1998,IsraelA&A1999}.}. 

\begin{figure*}[ht]
\sidecaption
\includegraphics[width=12cm,trim = 0cm 0.6cm 0cm 0.6cm]{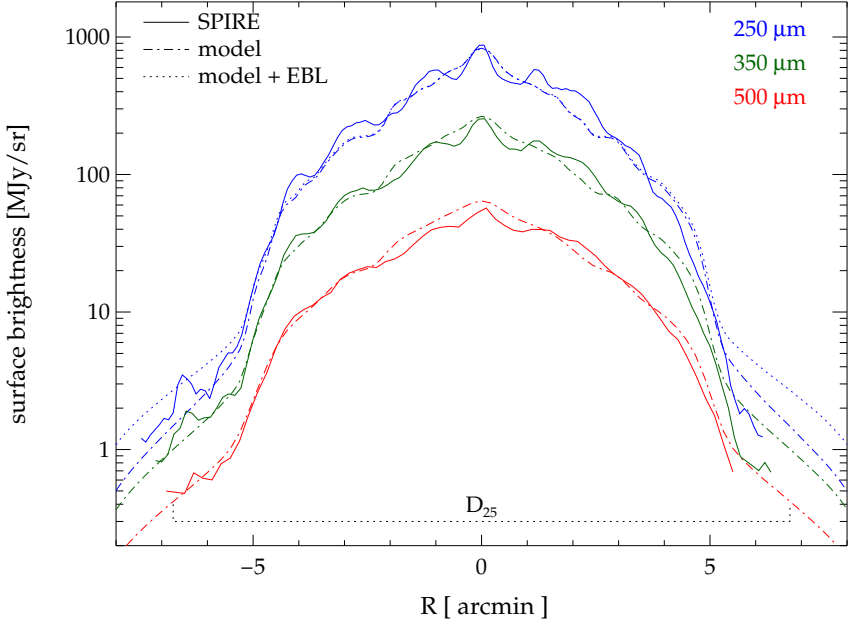}
\caption{Surface brightness profiles along the major axis from
SPIRE images (solid lines) and the corresponding models by \citet{BianchiA&A2008} 
(dot-dashed lines) at 250 (top, blue, lines), 350 (middle, green, lines), and 
500$\mu$m (bottom, red, lines).  We also show the profile
for the model that includes dust heating from the EBL (dotted line,
for 250$\mu$m only). For each band, profiles are averaged
over two beams perpendicular to the major axis. They are shown
for flux densities $\ge$ 1.2, 0.75, and 0.45 MJy sr$^{-1}$ for 250, 350, and 500$\mu$m,
respectively, corresponding
to S/N $>$ 3 for the average. Negative distances from the centre refer 
to the SW end of the galaxy. 
The optical size of the galaxy ($D_{25} =13\farcm 5$; 
\citealt{RC3}) is indicated. See text for other details.}
\label{figrad}
\end{figure*}

Major axis profiles are shown in Fig.~\ref{figrad}. To increase the
signal-to-noise ratio, they were averaged over two beams across the plane.
Residual sky gradients were estimated from two parallel strips on both 
sides of the galactic plane, well beyond the emission 
perpendicular to the major axis. They were found to be smaller than
0.5-$\sigma$ in all bands, so they do not affect the profiles significantly.
Profiles were extracted using a position angle of $22.9^\circ$,
which was determined with the fitting procedure of \citet{BianchiA&A2007}. 
The inner profile of the galaxy can be roughly described as the edge-on projection
of an exponential disk, with radial scalelength increasing from 4.6 kpc
at 250$\mu$m, to 5.1 kpc at 350$\mu$m, and to 5.2 kpc at 500$\mu$m.
(We assumed a distance D=9.5 Mpc; \citealt{VanDerKruitA&A1981}.) A break 
occurs in all bands at about 12 kpc from the centre (4$\farcm$4),
after which the profiles become steeper (with a radial scalelength of
about 1 kpc). Such radial breaks are analogous to what is found in 
most stellar disks \citep{PohlenA&A2006} and have already been observed 
with SPIRE \citep{PohlenA&A2010}. Though the change in slope in NGC~891
is quoted as appearing at about 6$\arcmin$
\citep[coming to a sharp truncation at 7$\farcm$5; ][]{VanDerKruitA&A1981},
the optical images used in \citet{XilourisSub1998} indicate
a break similar to the one detected for dust. 
SPIRE clearly improves over previous submm observations. SCUBA
was able to detect the dust emission at 850$\mu$m 
within the radial break only \citep{AltonApJL1998}. The 850$\mu$m morphology  
for the inner region is very similar to SPIRE images, and
a radial scalelength of 5.3 kpc was fitted to the profile
\citep{AltonSub1999}, close to what we find at 500$\mu$m.

After the break, the profile on the NE side of the galaxy can only be traced 
up to 6$\arcmin$, or 90\% of the optical size (assuming $D_{25} =13\farcm5$ = 37 kpc; 
\citealt{RC3}). On the SW end, instead, significant emission is seen in
excess of the break, up to the optical size and beyond. The excess
has already been detected in ISOPHOT images at 170 and 200$\mu$m
\citep{PopescuA&A2003b}; it corresponds to the SW extension 
of the asymmetric HI disk, as can be seen by comparing
the 3-$\sigma$ SW appendage of Fig.~\ref{figmap} 
with the 10$^{21}$ cm$^{-2}$ HI column density
contour at a comparable resolution in Fig.~1 of \citet{OosterlooAJ2007}. 
{
From the data, we derived a dust mass for this feature of $6\times10^5$ M\sun\ 
(though with an uncertainty of almost a factor two, mainly because of
the difficulty deriving temperatures using only SPIRE fluxes).
The HI mass corresponding to the SPIRE detection
(estimated from the column density levels
in \citealt{OosterlooAJ2007}) is about $10^8$ M\sun, resulting
in a dust-to-gas mass ratio of 0.006, similar to what is found for the
inner disks of other galaxies and of the Milky Way (see, e.g., 
\citealt{DraineApJ2007}). This confirms that the gas 
is not pristine, and it supports the view that the lopsidedness
of NGC~891 originated from the perturbation of the galactic disk
during a fly-by of the nearby companion UGC 1807 \citep{MapelliMNRAS2008}.
The HI mass for the whole SW disk extension is $2.5 \times 10^8$ M\sun.
Assuming the same dust-to-gas mass ratio, we estimated a 
total dust mass of $1.5\times10^6$ M\sun\ for the region.
(A similar result was found by \citealt{PopescuA&A2003b}.)

Previous radiative transfer models of NGC~891 underpredict the 
new submm SED. For instance, the 250$\mu$m flux density
predicted by \citet{BianchiA&A2008} is about 45\% lower
than observed (see Fig.~\ref{figsed}, where we also show the similar,
but independent, model of \citealt{PopescuA&A2011}). We revised the
model to describe 
the submm SED and the radial profiles observed by SPIRE better. Following the 
procedure of \citet{BianchiA&A2008}, we increased the mass in the 
{\em clumpy dust disk} (from 0.5 to 1.1$\times10^8$  M\sun, making up
70\% of the total dust mass)
and fine-tuned the other parameters (mainly, 
we raised the galaxy intrinsic luminosity from 7.8 to 8.5 
$\times10^8$ L\sun) to reproduce the stellar SED as well. The 
SED of the new model is shown in Fig.~\ref{figsed}. It is only marginally
consistent with the 100$\mu$m IRAS flux density, and it does not reproduce
the 170 and 200$\mu$m flux densities from ISO.
With the FIR/submm emissivity of the adopted dust
grain model \citep{DraineApJ2007b}, and for the interstellar
radiation fields implied by the relative geometries of dust and stars
assumed within the radiative transfer code TRADING, it was not possible 
to find a model able to describe all datapoints.

Modelled major axis profiles (Fig.~\ref{figrad}) were
obtained by convolving simulated images in the three bands
with the respective empirical PSFs, regridding them on the chosen
pixel size and averaging the profile as for real data.
For a match of the broad surface brightness distribution
along the major axis (as well as of the SED), we had to use
the same radial scalelength (5.7 kpc) both for the
{\em clumpy dust disk} and for the stellar disk. The second result 
is puzzling, since most of the observed stellar emission (and a good
fraction of the absorbed energy in the current model) comes from 
the NIR, {for which the radiative transfer fits found a 
smaller stellar radial scalelength \citep[4 kpc;][]{XilourisSub1998}}. 
Also, as already noted in
\citet{BianchiA&A2008}, emission from the clumpy dust disk appears
smoother than the real data, revealing that dust inhomogeneities
on a larger scale than molecular clouds (i.e. rings, spiral arms) 
should be considered in models.

The break in the profile is reproduced in the model by a sharp cut of 
both the clumpy dust disk and the stellar disk at 13 kpc from the centre
(4.7$\arcmin$ in Fig.~\ref{figrad})\footnote{Conversely, the 
break could be produced by truncating the
{\em clumpy dust disk} only, if the model did not include a 
{\em diffuse dust disk}.
}. Beyond the break, the modelled emission
is only due to the diffuse dust disk derived from optical observations.
The mean dust temperature in this disk is low ($10 < T < 13 K$) because
the heating sources (mainly stars at the edge of the disk; the bulge
contribution is negligible) are no longer mixed with dust.
The predicted surface brightness matches the observations
only on the SW, where emission comes from the excess 
previously described.
The axisymmetric diffuse dust disk derived
by \citet{XilourisSub1998} could thus mimic the contribution
to extinction of the (asymmetric) extended HI component.
However, it is to be noted that the model predicts a lower
temperature than what is derived in the SW appendage (15K),
and the mass of the {\em diffuse dust disk} beyond 13 kpc
(8$\times10^6$ M\sun) is $5\times$ higher than what is estimated
from the HI data, which is too high a factor to be explained by
the uncertainties in the temperature and in the various model 
assumptions. Also, model predictions are likely lower limits to the 
real emission from a diffuse dust disk. If stars
extend more than what is needed to fit the radial break, the
emission will be higher. Instead, if grains are isolated from stars, 
the extragalactic background light (EBL) might contribute significantly.
The dotted line in Fig.~\ref{figrad} shows the 
effect on the 250$\mu$m profile of the EBL P1.0
model of \citet{AharonianNature2006}. Emission from the
diffuse dust disk in this region of the model could rise by a factor 
two\footnote{The temperature of dust heated by the EBL only
is about 11 K, adopting mean dust properties from \citet{DraineApJ2007b}.}.
}

\begin{figure}
\centering
\resizebox{\hsize}{!}{\includegraphics[trim = 0cm 0.6cm 0cm 0.5cm]{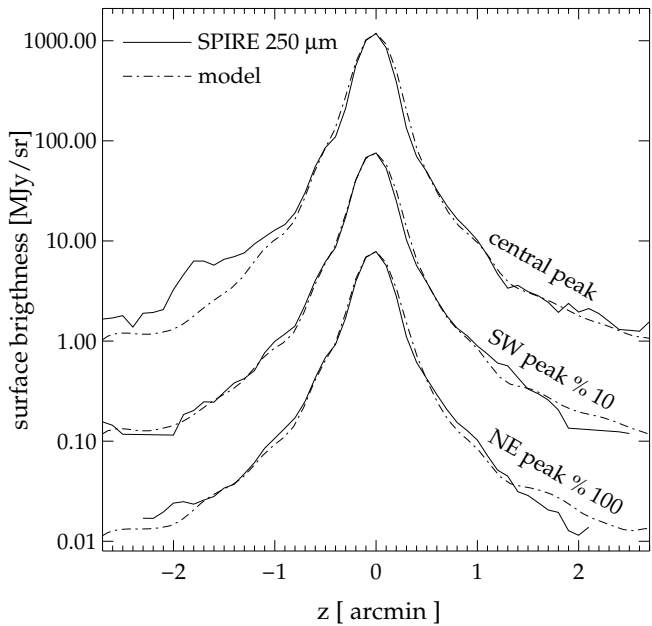}}
\caption{Data and (scaled) model profiles at 250$\mu$m perpendicular 
to the major axis through the centre of the galaxy and 
the two secondary peaks. The profile through the SW peak has been 
divided by 10, that in the NE by 100).  Negative heights refer 
to the NW side of the disk.}
\label{figver}
\end{figure}

Finally, Fig.~\ref{figver} shows the 250$\mu$m profiles
perpendicular to the major axis, for three cuts passing through
the galaxy centre and the two secondary peaks at about 1$\arcmin$ 
from the centre on the NE and SW.
Data were averaged over two beams perpendicular to the cut,
and a residual background was subtracted from regions beyond the
emission. Model profiles were scaled to match 
the data on the major axis. Emission above
the plane is almost entirely dominated by the PSF Airy rings
picking up light from the major axis (because the adopted dust vertical 
scalelength, 0.2~kpc or 4$\arcsec$, is much smaller than
the main beam). Results are similar in the other SPIRE bands.
Excess emission is only seen on the NW side 
of the minor axis profile, in coincidence with the
background X-ray source CXO~J022224.4+422138, which 
contributes to most of the excess; however, part of it could
come from another background source aligned with the minor axis
or from a filamentary structure protruding from the galactic centre.
Excluding this feature, SPIRE images do not show evidence of any 
diffuse halo emission. This is  at odds with the previous
Spitzer observations of \citet{BurgdorfApJ2007}, which detected 
MIR emission extending up to 5 kpc from the plane.  At 22$\mu$m,
a surface brightness of 0.1-0.2 MJy/sr is found at 1-1\farcm4
along the minor axis. Instead, at 250$\mu$m, residual emission from
the halo is less than 1 MJy/sr. Such MIR/submm flux density 
ratios can be produced by typical Milky Way dust, if the 
interstellar radiation field is 10$\times$ higher than the local 
(see, e.g., Fig. 13 in \citealt{DraineApJ2007b}). These heating
conditions might be compatible with the hypothesis that
the detected MIR radiation come from
dust surrounding halo AGB stars \citep{BurgdorfApJ2007}.

\section{Summary \& conclusions}

We have analysed SPIRE observations of the edge-on spiral galaxy NGC~891
and examined both their contribution to the galaxy's SED and the
galaxy's morphological properties at these wavelengths. 
Emission within the central 4$\farcm$4 broadly follows a
radial exponential distribution with a scalelength similar
to those derived for the stars in the surface brightness 
fits of \citet{XilourisSub1998}. Emission above
the galactic plane can be explained with radiation from
the galactic plane picked up by the PSF wings. The 
distribution of dust along the vertical direction is 
thus thinner than the SPIRE PSFs, and compatible with
the vertical scalelength derived by \citet{XilourisSub1998}.

Beyond 4$\farcm$4 from the centre, the radial profile shows a 
break and becomes steeper. On the NE side, the submm disk comes 
to an end within the optical size of the galaxy. Instead, on 
the SW side the profile decline is interrupted at about the
optical radius, where significant emission is seen in all bands. 
With their high resolution, SPIRE observations confirm the 
excess emission associated with the asymmetric HI disk, already 
found in deep ISOPHOT observations by \citet{PopescuA&A2003b}.

The emission within the central 5$\arcmin$ can be attributed
to the {\em clumpy dust disk} required by radiative transfer models
to fit the FIR/submm SED of the galaxy \citep{BianchiA&A2008}. 
The HI-associated dust emission in the SW instead has
surface brightness levels that are marginally consistent with 
those predicted for the 
{\em diffuse dust disk}, an important dust component (of similar
mass to that of the clumpy dust disk) responsible for the optical/NIR 
extinction lane \citep{XilourisSub1998}. 
Since extended {\em diffuse dust disks} are routinely found in optical/NIR 
radiative transfer fits of edge-on galaxies, we hope that further
Herschel observations of these objects (as those planned within the 
approved projects HEROES and NHEMESES), together with radiative 
transfer models and deep observations of their atomic gas disks, will 
help for understanding if the diffuse dust disk is indeed a component 
associated with the atomic gas and is a feature common to all spiral galaxies.

\bibliographystyle{aa}
\bibliography{/Users/sbianchi/Documents/tex/DUST}

\end{document}